\documentclass{article}
\usepackage{header}
\usepackage{flushend}

\title{JOINT DEREVERBERATION AND SEPARATION with ITERATIVE SOURCE STEERING}

\name{%
  Taishi Nakashima$^{\diamondsuit, \clubsuit}$\thanks{This work was done while Taishi Nakashima was an intern at LINE Corporation.},
  Robin Scheibler$^{\clubsuit}$,
  Masahito Togami$^{\clubsuit}$,
  Nobutaka Ono$^{\diamondsuit}$
}
\address{%
  $^{\diamondsuit}$ Tokyo Metropolitan University, Tokyo, Japan.\\
  $^{\clubsuit}$ LINE Corporation, Tokyo, Japan.
}

\usepackage{textcomp}
\usepackage[absolute,showboxes]{textpos}

\TPMargin{5pt}

\newcommand{\copyrightstatement}{
	\begin{textblock}{0.84}(0.08,0.01)    % tweak here: {box width}(leftposition, rightposition)
		\noindent
		\footnotesize
		\textcopyright~2021~IEEE.
    Personal use of this material is permitted.
		Permission from IEEE must be obtained for all other uses,
    in any current or future media,
    including reprinting/republishing this material for advertising or promotional purposes,
    creating new collective works,
    for resale or redistribution to servers or lists,
    or reuse of any copyrighted component of this work in other works.
	\end{textblock}
}
\begin{document}
\copyrightstatement
\ninept
\maketitle
\bstctlcite{IEEE:BSTcontrol}

\begin{abstract}
We propose a new algorithm for joint dereverberation and blind source separation (DR-BSS).
Our work builds upon the IRLMA-T framework that applies a unified filter combining dereverberation and separation.
One drawback of this framework is that it requires several matrix inversions, an operation inherently costly and with potential stability issues.
We leverage the recently introduced iterative source steering (ISS) updates to propose two algorithms mitigating this issue.
Albeit derived from first principles, the first algorithm turns out to be a natural combination of weighted prediction error (WPE) dereverberation and ISS-based BSS, applied alternatingly.
In this case, we manage to reduce the number of matrix inversion to only one per iteration and source.
The second algorithm updates the ILRMA-T matrix using only sequential ISS updates requiring no matrix inversion at all.
Its implementation is straightforward and memory efficient.
Numerical experiments demonstrate that both methods achieve the same final performance as ILRMA-T in terms of several relevant objective metrics.
In the important case of two sources, the number of iterations required is also similar.
\end{abstract}


\section{Introduction}

Speech signals recorded by a microphone are routinely contaminated by reverberation and interference.
Blind source separation (BSS)~\cite{Makino:2007:BSS,Makino:2018:ASS,Sawada:2019:APSIPATSIP}, e.g., independent component analysis (ICA)~\cite{Comon:1994:ElsevierSP}, independent vector analysis (IVA)~\cite{Kim:2006:ASLP,Hiroe:2006:ICA,Ono:2010:LVAICA,Ono:2011:WASPAA}, and dereverberation (DR)~\cite{Naylor:2010:SD} techniques, e.g., weighted prediction error (WPE)~\cite{Nakatani:2010:ASLP}, are all countermeasures that have been proposed to recover the speech quality required for communication, speech diarization, and automatic speech recognition (ASR) systems.
Historically, DR and BSS have evolved separately, and their joint optimization has not yet matured.
Joint optimization is highly desirable to realize DR and BSS in the same framework (DR-BSS) as it typically leads to higher speech quality.

DR-BSS algorithms have been actively studied since WPE~\cite{Nakatani:2010:ASLP} was introduced~\cite{Yoshioka:2011:ASLP, Togami:2013:ASLP,Kagami:2018:ICASSP,Ikeshita:2019:EUSIPCO,Ikeshita:2019:WASPAA,Togami:2020:ICASSP}.
A popular approach is to combine WPE~\cite{Nakatani:2010:ASLP} with a BSS algorithm such as Independent Low-Rank Matrix Analysis (ILRMA)~\cite{ Kitamura:2016:ASLP}.
Early studies~\cite{Togami:2013:ASLP,Kagami:2018:ICASSP} use separate DR and BSS filters.
However, computational cost of these approaches is very high due to the necessity of computing the inverse of a large matrix whose dimension is the product of the \textit{square} of the numbers of microphones with the number of taps of the DR filter.
The recently proposed ILRMA-T~\cite{Ikeshita:2019:EUSIPCO,Ikeshita:2019:WASPAA} overcomes this difficulty by introducing a unified filter combining the DR and BSS filter.
Nevertheless, ILRMA-T still requires to invert two matrices per source and iteration.
Because DR-BSS algorithms are typically needed in edge and embedded devices, where computational power is at a premium, inverse matrix computations are best avoided.

ILRMA-T derives the update equations for its DR-BSS matrix from the iterative projection (IP) rules of BSS~\cite{Ono:2011:WASPAA}.
In the BSS context, some of the authors have proposed iterative source steering (ISS), an alternative to IP that is more computationally efficient and does not require matrix inversion~\cite{Scheibler:2020:ICASSP}.
Thus, the ISS based approach is more stable than the IP one.
To the best of our knowledge, ISS based DR-BSS has not been studied yet.

In this paper, we propose a joint optimization framework for DR-BSS with ISS~\cite{Scheibler:2020:ICASSP}.
The proposed method optimizes the same cost function as ILRMA-T, but using the ISS updates\@.
Thus, we call it ILRMA-T-ISS\@.
Two variants of ILRMA-T-ISS are proposed.
The first one is obtained by updating all the weights in the DR-BSS matrix corresponding to dereverberation in a single step, and apply ISS for the rest.
The resulting algorithm turns out to be a natural combination of WPE and ISS, with their respective updates applied alternatingly.
We call this algorithm ILRMA-T-ISS-JOINT\@.
ILRMA-T-ISS-JOINT reduces the number of matrix inversions to only one per iteration and source.
The second variant, ILRMA-T-ISS-SEQ, applies sequential ISS updates to the whole matrix.
This has the happy consequence that not a single matrix inversion is required.
One practical consequence is that its implementation is straightforward and no external linear algebra library is needed.
These properties are all highly desirable in edge and embedded systems.
We conduct numerical experiments to confirm the efficacy of the proposed method in noisy reverberant environment with multiple speech sources.
We confirm that separation and dereverberation performance are on par with ILRMA-T-IP, even without the matrix inversions.

\section{Background}
\subsection{Signal model and notation}
Let $\NumSource$ and $\NumMic$ be the numbers of sources and microphones, respectively.
Henceforth, we consider the determined case, $\NumSource = \NumMic$.
We use the short-term Fourier transform (STFT) representation of microphone input signals. The microphone input signal is modeled as the following convolutive mixture:
\begin{align}
  \observ_{\IdxFreq, \IdxTime} &=
  \sum_{d=0}^{D-1}
  \mixMatrix_{\IdxFreq,d} \bm{s}_{\IdxFreq,\IdxTime-d} \in \C^{\NumSource}, \label{input_model}
\end{align}
where
$\IdxFreq \in \{1, \dots, \NumFreq\}$
and
$\IdxTime \in \{1, \dots, \NumTime\}$
are the frequency bin and the time frame indices, respectively,
$\mixMatrix_{\IdxFreq,d}$ is the mixing matrix with $(\mixMatrix_{\IdxFreq,d})_{\IdxSource, \IdxMic} = a_{\IdxSource, \IdxMic, \IdxFreq,d}$,
$\source$ is the source signal,
and
$\IdxSource = \{1, \dots, \NumSource\}$ is the source channel index.

In the rest of the manuscript,
$^{\top}, ^{\hermite}$, and $\det$
denote the transpose, Hermitian transpose, and determinant of a vector/matrix, respectively.
We denote the $\IdxSource$th canonical basis vector by $\eyeVec_{\IdxSource}$, an all zero vector $\bm{0}$, and the identity matrix by $\bm{I}$.

\subsection{Dereverberation based on Weighted prediction error (WPE)}
WPE~\cite{Nakatani:2010:ASLP} is a popular approach for DR\@.
In WPE,~\eqref{input_model} is converted to the following auto-regressive (AR) model:
\begin{align}
      \observ_{\IdxFreq, \IdxTime} &=
  \sum_{\tau=0}^{L-1}
  \bm{Z}_{\IdxFreq,\tau} \observ_{\IdxFreq, \IdxTime-\Delta-\tau}, \label{ar_model}
\end{align}
where $\bm{Z}$ is a matrix which contains the AR coefficients, and
$L$ is the tap-length of the AR model. The WPE assumes that there is only one speech source, and $\bm{Z}$ is optimized with the time-varying variance of the speech source $r_{f,t}$ as follows:
\begin{align}
  % \overline{\bm{Z}}_{f}= \left( \sum_{t} \frac{ \observExtExt_{\IdxFreq, \IdxTime} \observExtExtH_{\IdxFreq, \IdxTime}}{r_{f,t}} \right) \left(\sum_{t} \frac{\observExtExt_{\IdxFreq, \IdxTime} \observExtExtH_{\IdxFreq, \IdxTime}}{r_{f,t}}  \right)^{-1}, \label{wpe_update}
  \bar{\bm{Z}}_{f} =
  \left( \sum _{t} \frac{\observExtExt _{f,t} \observExtExtH _{f,t}}{r _{f,t}} \right) ^{-1}
  \begin{bmatrix}
    \sum _{t} \frac{\observExtExt _{f,t} x _{1,f,t} ^{\ast}}{r _{f,t}}
      & \dots &
    \sum _{t} \frac{\observExtExt _{f,t} x _{N,t} ^{\ast}}{r _{f,t}}
  \end{bmatrix} ^\top
\end{align}
where $\observExtExt_{\IdxFreq, \IdxTime} =
    \begin{bmatrix}
      \observT_{\IdxFreq, \IdxTime-\Delay}  &
      \cdots &
      \observT_{\IdxFreq, \IdxTime-\Delay-\NumTap+1}
    \end{bmatrix} ^{\top} \in \C^{\NumSource\NumTap}$,
$\Delay$ is the delay,
and
$\overline{\bm{Z}}_{f}=\begin{bmatrix}
 \bm{Z}_{\IdxFreq,0} &
        \cdots &
        \bm{Z}_{\IdxFreq,L-1}
\end{bmatrix}$.
The dereverberated signal $\bm{z}_{f,t}$ is obtained as $\bm{z}_{f,t}=\observ_{\IdxFreq, \IdxTime} - \overline{\bm{Z}}_{f} \observExtExt_{\IdxFreq, \IdxTime}$.
Then, we update $r_{f,t}=\frac{\lVert \bm{z}_{f,t}\rVert^{2}}{M}$.
Thus, $\overline{\bm{Z}}_{\IdxFreq}$ and $r_{f,t}$ are updated in an iterative manner.

\subsection{Joint dereverberation and separation}
Cascade connection of the WPE and the BSS is not optimum because the WPE assumes that there is only one source.
In~\cite{Togami:2013:ASLP,Kagami:2018:ICASSP}, joint optimization of the WPE and the BSS  is performed by using a WPE filter followed by a BSS filter.
The output signal is obtained as $
\bm{y}_{f,t}=\bm{W}_{f}\left(\observ_{\IdxFreq, \IdxTime} - \overline{\bm{Z}}_{f} \observExtExt_{\IdxFreq, \IdxTime} \right)$.
A determined approach is proposed in~\cite{Kagami:2018:ICASSP} for optimization of $\bm{W}_{f}$ and $\overline{\bm{Z}}_{f}$ sequentially,
such that the separated signal $\bm{y}_{f,t}$ is the maximum likelihood estimator of $\source$ under the assumptions
\begin{enumerate}
  \item the sources are statistically independent,
  \item a source signal at each time-frequency bin belongs to a complex Gaussian distribution: \\
    $
      p(y_{\IdxSource, \IdxFreq, \IdxTime}) =
        \frac{1}{\pi \variance_{\IdxSource, \IdxFreq, \IdxTime}}
        \exp \left(- \frac{\lvert y_{\IdxSource, \IdxFreq, \IdxTime} \rvert ^2}{\variance_{\IdxSource, \IdxFreq, \IdxTime}} \right)$,
    where $y_{\IdxSource, \IdxFreq, \IdxTime}$ is the $\IdxSource$th element of $\estimate$ and
    $\variance_{\IdxSource, \IdxFreq, \IdxTime}$ is the time-varying variance of the $\IdxSource$th source,
  \item $\variance_{\IdxSource, \IdxFreq, \IdxTime}$ is modeled as $
      \variance_{\IdxSource, \IdxFreq, \IdxTime} = \sum _{\IdxBasis = 1} ^{\NumBasis} \NMFbasis_{\IdxSource, \IdxBasis, \IdxFreq} \NMFactivation_{\IdxSource, \IdxTime, \IdxBasis}$,
    where $\NumBasis$ is the number of basis vectors, $\NMFbasis_{\IdxSource, \IdxBasis, \IdxFreq} \geq 0$ is the basis coefficient of the $\IdxSource$th component,
    and $\NMFactivation_{\IdxSource, \IdxTime, \IdxBasis} \geq 0$ is the time-varying activity of the $\IdxSource$th component.
\end{enumerate}
Parameters are updated to maximize the following negative log-likelihood function $\cost$:
\begin{align}
  \cost &=
  \sum _{\IdxFreq, \IdxTime} \Biggl[
      - 2 \log \lvert \det \demixMatrix_{\IdxFreq} \rvert
      + \sum _{\IdxSource} \left(
        \frac{{|y_{n,f,t}|}^{2}}{\variance _{\IdxSource,\IdxFreq,\IdxTime}}
        + \log \variance _{\IdxSource,\IdxFreq,\IdxTime}
      \right)
  \Biggr].
  \label{eq:cost:kagami}
\end{align}
% where
% $\displaystyle \covariance _{\IdxSource, \IdxFreq} =\frac{1}{\NumTime} \sum _{\IdxTime} \frac{
%  \left(\observ_{\IdxFreq, \IdxTime} - \overline{\bm{Z}}_{f} \observExtExt_{\IdxFreq, \IdxTime} \right)
%  \left(\observ_{\IdxFreq, \IdxTime} - \overline{\bm{Z}}_{f} \observExtExt_{\IdxFreq, \IdxTime} \right)^\hermite}{\variance _{\IdxSource, \IdxFreq, \IdxTime}}
%  \in \C^{\NumSource\times\NumSource}$
% weighted covariance matrix of the dereverberated signal.
The IP based parameter optimization~\cite{Ono:2011:WASPAA} can be straightforwardly applied for optimization of $\bm{W}_{f}$.
Non-negative matrix factorization (NMF) is used to update $\NMFbasis_{\IdxSource, \IdxBasis, \IdxFreq}$  and $\NMFactivation_{\IdxSource, \IdxTime, \IdxBasis}$~\cite{Kitamura:2016:ASLP}.
The optimal $\overline{\bm{Z}}_{f}$ is also obtained straightforwardly by minimizing $\cost$.
However, when $\overline{\bm{Z}}_{f}$  is updated, it is necessary to calculate the inverse matrix of a large-scale matrix whose dimension is proportional to $M^2 L$.
Thus, computational cost is quite high.

As an alternative, ILRMA-T~\cite{Ikeshita:2019:EUSIPCO,Ikeshita:2019:WASPAA} has been proposed.
ILRMA-T combines WPE and ILRMA~\cite{Kitamura:2016:ASLP} for joint dereverbertion and separation.
In ILRMA-T, the output signal is obtained by a unified filter $\bm{P}_{f}$ as $
\bm{y}_{f,t}=\bm{P}_{f}\observExt_{\IdxFreq, \IdxTime}$,
where
$\observExt_{\IdxFreq, \IdxTime} =
    \begin{bmatrix}
      \observT_{\IdxFreq, \IdxTime} &
      \observExtExtT_{\IdxFreq, \IdxTime}
    \end{bmatrix} ^{\top} \in \C^{\NumSource(\NumTap+1)}$
and
$\bm{P}_{f} =
\bm{W}_{f}
\begin{bmatrix}
  \bm{I} &
  -\overline{\bm{Z}}_{f}
\end{bmatrix}$.

The cost function of ILRMA-T is equivalent to~\eqref{eq:cost:kagami}, that is,
\begin{align}
  \cost &=
    \sum _{\IdxFreq} \left[
      - 2 \log \lvert \det \demixMatrix_{\IdxFreq} \rvert
      + \sum _{\IdxSource} \derevVectorH_{\IdxSource, \IdxFreq} \covariance_{\IdxSource, \IdxFreq} \derevVector_{\IdxSource, \IdxFreq}
    \right],
\end{align}
where
$\displaystyle \covariance _{\IdxSource, \IdxFreq} =
    \frac{1}{\NumTime}
    \sum _{\IdxTime} \frac{\observExt _{\IdxFreq, \IdxTime} \observExtH _{\IdxFreq, \IdxTime}}{\variance _{\IdxSource, \IdxFreq, \IdxTime}}
    \in \C^{\NumSource(\NumTap+1)\times\NumSource(\NumTap+1)}$ is the weighted covariance matrix of $\observExt_{f,t}$.

Instead of optimizing $\bm{W}_{f}$ and $\bm{Z}_{f}$ sequentially,  ILRMA-T optimizes each row vector of $\bm{P}_{f}$ sequentially based on IP~\cite{Ono:2011:WASPAA}.
The filter to separate and dereverberate the $\IdxSource$th source is defined as  $\derevVectorH_{\IdxSource, \IdxFreq}$, i.e., the $n$th row vector of $\bm{P}_{f}$.
It is updated as follows:
\begin{align}
\derevVector _{\IdxSource, \IdxFreq} \leftarrow \frac{\bm{V}_{n,f}^{-1} \bm{a}_{n,f}}{\sqrt{ \bm{a}_{n,f}^{\hermite}\bm{V}_{n,f}^{-1} \bm{a}_{n,f}} }, \label{p_update}
\end{align}
where $
\bm{a}_{n,f}=\begin{pmatrix} \bm{W}_{f}^{-1} \bm{e}_{n}  \\ \bm{0} \end{pmatrix}$.
Thus, calculation of two types of inverse matrices are needed in the $\derevVector _{\IdxSource, \IdxFreq}$ update.
The updates of $\NMFbasis_{\IdxSource, \IdxBasis, \IdxFreq}$ and $\NMFactivation_{\IdxSource, \IdxTime, \IdxBasis}$ are those of NMF\@.
We call this algorithm ILRMA-T-IP\@.

%! TEX Root = ../paper_2021_ICASSP_Nakashima.tex
\section{Proposed method: ILRMA-T-ISS}
We propose a new DR-BSS method to reduce the number of inverse matrix computations.
The cost function is the same as that of the ILRMA-T, which is defined as
\begin{align}
  \cost &=
    \sum _{\IdxFreq} \left[
      - 2 \log \lvert \det \demixMatrix_{\IdxFreq} \rvert
      + \sum _{\IdxSource} \derevVectorExtH_{\IdxSource, \IdxFreq} \covariance_{\IdxSource, \IdxFreq} \derevVectorExt_{\IdxSource, \IdxFreq}
    \right],
    \label{eq:ilrma_t_cost}
\end{align}
where
\begin{itemize}
  \item $\derevMatrixExt_{\IdxFreq} =
    \begin{bmatrix}
      \derevMatrix_{\IdxFreq} \\
      \zeros_{\NumSource\NumTap\times\NumSource}\quad\eyeMat_{\NumSource\NumTap}
    \end{bmatrix} \in \C^{\NumSource(\NumTap+1)\times\NumSource(\NumTap+1)}$
  \item $\derevVectorExtH_{\IdxSource, \IdxFreq}$: $\IdxSource$th row vector of $\derevMatrixExt_{\IdxFreq}$.
\end{itemize}
Optimization of the parameters is done via ISS~\cite{Scheibler:2020:ICASSP}.
When $n \le N$,
ISS updates $\derevMatrixExt$ (the index of frequency bins omitted) like this,
\begin{align}
  \derevMatrixExt \gets \derevMatrixExt -
    \begin{bmatrix}
      \updateCoef_{1,\IdxSource} \\
      \vdots \\
      \updateCoef_{\NumSource,\IdxSource} \\
      \zeros _{\NumSource\NumTap\times 1}
    \end{bmatrix}
  \derevVectorExtH _{\IdxSource}.
\end{align}
This update rule is the same as that for BSS\@.
The minimization of~\eqref{eq:ilrma_t_cost} with respect to $\updateCoef_{m,\IdxSource}$ gives,
    \begin{align}
      \updateCoef _{\IdxMic, \IdxSource} &=
      \begin{cases}
        \dfrac{\derevVectorExtH _{\IdxMic} \covariance _{\IdxMic} \derevVectorExt _{\IdxSource}}
             {\derevVectorExtH _{\IdxSource} \covariance _{\IdxMic} \derevVectorExt _{\IdxSource}} & (\IdxMic \neq \IdxSource), \\
                                                                                                   & \\
        1 - (\derevVectorExtH _{\IdxSource} \covariance _{\IdxSource} \derevVectorExt _{\IdxSource})^{-\frac{1}{2}} & (\IdxMic = \IdxSource).
      \end{cases}\\
      & \qquad \forall 1 \leq \IdxMic \leq \NumSource
    \end{align}

For $n > N$, we propose two update rules, i.e., IRLMA-T-ISS-JOINT and ILRMA-T-ISS-SEQ\@.
These update rules correspond to the dereverberation part of the algorithm.

\subsection{ILRMA-T-ISS-JOINT}

We call the first update rule ILRMA-T-ISS-JOINT as it jointly updates $v_{m,n > N}$  in the following way,
\begin{align}
  \derevMatrixExt \gets \derevMatrixExt -
    \begin{bmatrix}
      \bm{v}_{1,n>N} \\
      \vdots \\
      \bm{v}_{\NumSource,n>N} \\
      \zeros _{\NumSource\NumTap\times\NumSource\NumTap}
    \end{bmatrix}
    \bm{G}_{n>N}^\hermite
  \nonumber
\end{align}
where $\bm{v}_{m,n>N} =\begin{bmatrix} \updateCoef _{\IdxMic, N+1} & \cdots & \updateCoef _{\IdxMic, N(L+1)}  \end{bmatrix}$ and
$\bm{G}_{n>N}=\begin{bmatrix} \derevVectorExt_{N+1} & \cdots &   \derevVectorExt_{N(L+1)} \end{bmatrix}$.
Minimization of~\eqref{eq:ilrma_t_cost} with respect to $\bm{v}_{m,\IdxSource > N}$ for $1\leq m \leq \NumSource$, yields,
    \begin{align}
      \bm{v}_{m,n>N} &=
      \left(\derevVectorExtH _{\IdxMic} \covariance _{\IdxMic}\bm{G}_{n>N} \right)
      \left( \bm{G}_{n>N}^\hermite \covariance _{\IdxMic}\bm{G}_{n>N} \right)^{-1},
    \end{align}
     which can be further expanded as follows,
    \begin{align}
\bm{v}_{m,n>N} = \left( \sum_{t} \frac{ y_{m, \IdxFreq, \IdxTime} \observExtExtH_{\IdxFreq, \IdxTime}}{r_{m,f,t}} \right) \left(\sum_{t} \frac{\observExtExt_{\IdxFreq, \IdxTime} \observExtExtH_{\IdxFreq, \IdxTime}}{r_{m,f,t}}  \right)^{-1}. \label{iss_wpe}
\end{align}
This equation is very similar to the update of the WPE filter by~\eqref{wpe_update}.
The latter is updated from the cross correlation between the current microphone input signal and the past microphone input signal.
On the other hand, in~\eqref{iss_wpe}, cross-correlation between the estimated output signal of the $m$th speech source and the past microphone input signal is calculated.
Thus, WPE based DR and ISS based BSS are naturally combined in this framework.
Moreover, it only requires inversion of one $NL\times NL$ matrix, in~\eqref{iss_wpe}, per iteration and source.

\subsection{ILRMA-T-ISS-SEQ}

We call the second update rule ILRMA-T-ISS-SEQ\@.
Instead of the joint update of $\bm{v}_{m,n>N}$, $v_{m,n}$ is updated for each $n>N$ sequentially as follows,
\begin{align}
  \derevMatrixExt \gets \derevMatrixExt -
    \begin{bmatrix}
      \updateCoef_{1,\IdxSource} \\
      \vdots \\
      \updateCoef_{\NumSource,\IdxSource} \\
      \zeros _{\NumSource\NumTap\times 1}
    \end{bmatrix}
  \derevVectorExtH _{\IdxSource}. \nonumber
\end{align}
Minimization of~\eqref{eq:ilrma_t_cost} with respect to $\updateCoef_{m,N}$ gives,
    \begin{align}
      v_{m,n} &= \frac{ \derevVectorExtH _{\IdxMic} \covariance _{\IdxMic}\bm{g}_{n} }{\bm{g}_{n}^\hermite \covariance _{\IdxMic}\bm{g}_{n}},
      \quad \forall 1 \leq \IdxMic \leq \NumSource \nonumber.
    \end{align}
This can be further expanded as,
\begin{align}
 v_{m,n} =
 \frac{
   \sum_{t} \frac
   {y_{m, \IdxFreq, \IdxTime} \tilde{x}_{n, \IdxFreq, \IdxTime}^{*}}
   {r_{m,f,t}}
 }{
   \sum_{t} \frac
     {\tilde{x}_{n, \IdxFreq, \IdxTime} \tilde{x}_{n, \IdxFreq, \IdxTime}^{*}}
     {r_{m,f,t}}
 },
\end{align}
where
$\tilde{x}_{n, \IdxFreq, \IdxTime}$
is the $n$th element of $\observExt _{\IdxFreq, \IdxTime}$.
It is shown that inverse calculation is completely unnecessary  in  ILRMA-T-ISS-SEQ\@.

\newcommand{\pyroom}{\texttt{pyroom\-acoustics}}
\intervalconfig{soft open fences}
\section{Experiment}
\begin{figure*}[htb]
  \centering
  \includegraphics[width=\textwidth]{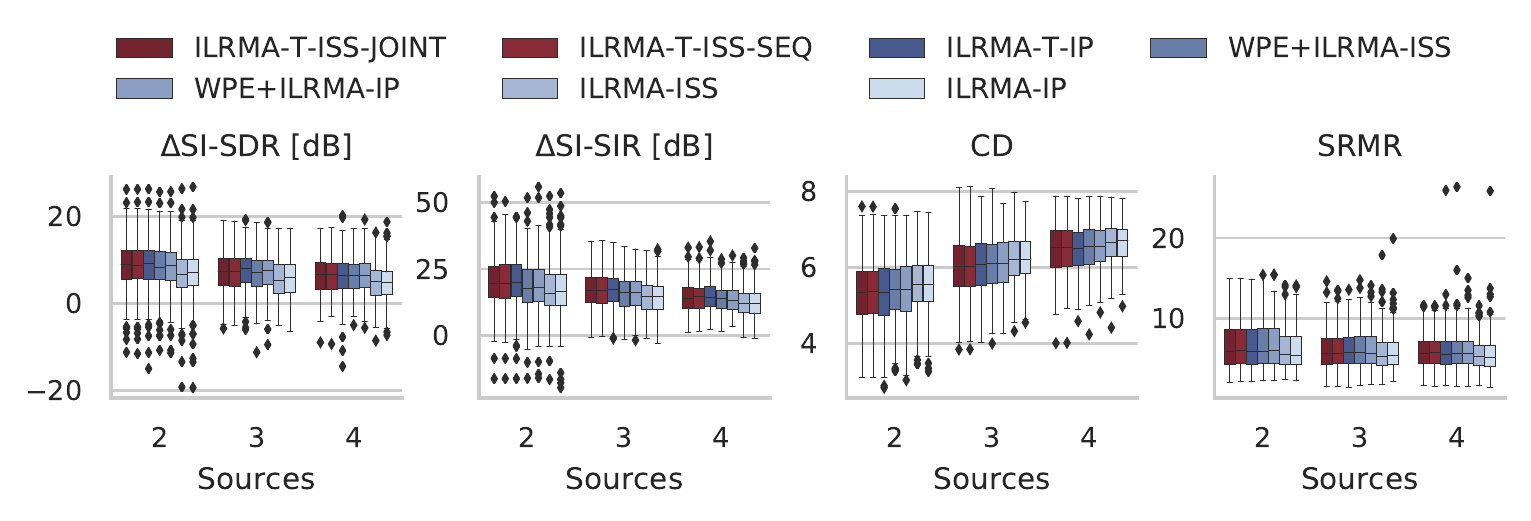}
  \caption{Average SI-SDR improvements, SI-SIR improvements, CD, and SRMR after 100 iterations. Higher is better for all metrics, except CD, for which lower is better.}%
  \label{fig:box}
\end{figure*}
\begin{figure*}[htb]
  \centering
  \includegraphics[width=\textwidth]{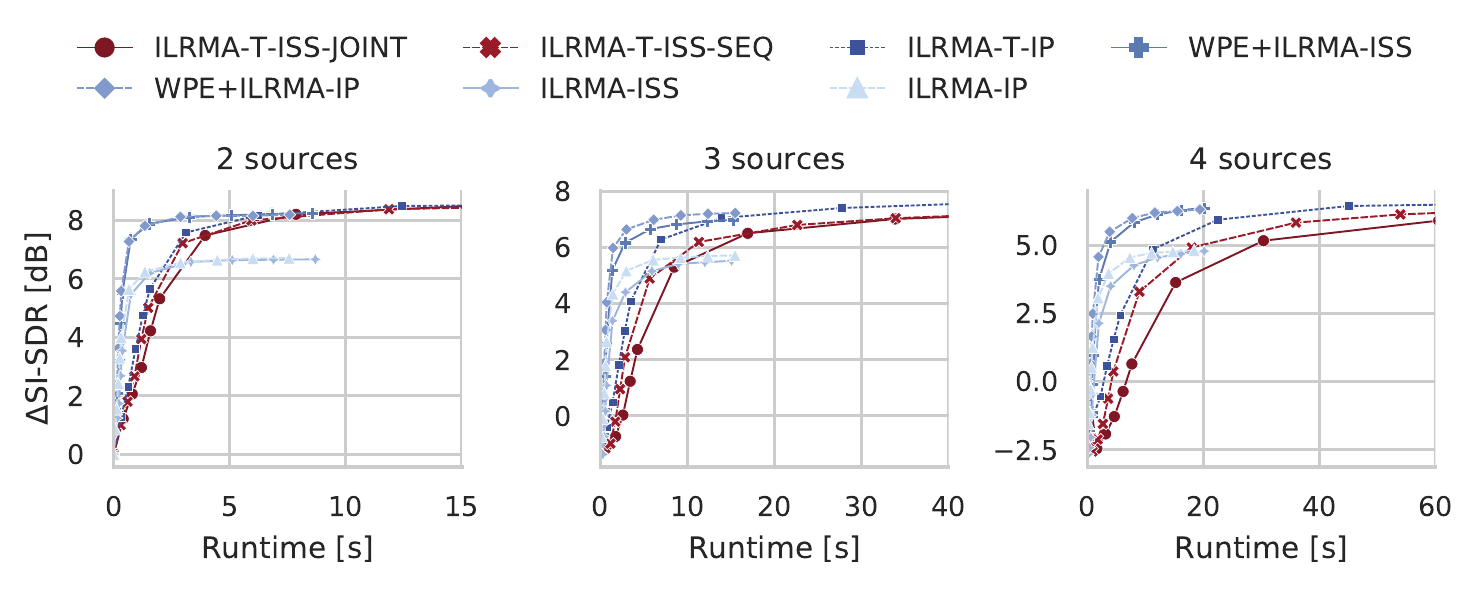}
  \caption{Convergence curves of average SI-SDR improvements for varying number of sources in 100 iterations.}%
  \label{fig:sdr:line}
\end{figure*}

\subsection{Setup}
We use speech sources from the WSJ corpus~\cite{NIST:1994:LDC} for evaluation.
To make the reverberant mixtures, we perform room simulations with the \pyroom{} Python package~\cite{Scheibler:2018:ICASSP} in random rectangular rooms with
walls between \SI{5}{\meter} and \SI{10}{\meter} length,
ceiling between \SI{3}{\meter} and \SI{4}{\meter} high.
Simulated reverberation times range from \SI{200}{\milli\second} to \SI{600}{\milli\second}.
The microphone array is circular, with a radius between \SI{0.075}{\meter} and \SI{0.125}{\meter}, such that the spacing is at least \SI{0.05}{\meter}.
The horizontal location of the microphone array and the speech sources is randomly chosen at least \SI{0.2}{\meter} away from the center of the rooms and at least \SI{1.5}{\meter} away from the center of the microphone array, respectively.
The vertical location of the microphone array and the sources ranges from \SIrange{1.0}{2.0}{\meter} and from \SI{1.5}{\meter} and \SI{2.0}{\meter} high, respectively.
The distance between the sources is randomly set to be at least \SI{1.0}{\meter}.
We add background noise selected from the CHiME3 dataset~\cite{Barker:2017:CSL} to each simulated signal.
The source signals are normalized to have unit power at the first microphone.
Then we define signal-to-noise ratio $\mathsf{SNR} = \NumSource / {\sigma^2}$, where $\sigma^2$ is the variance of uncorrelated white noise at the microphones.
The $\mathsf{SNR}$ ranges from \SIrange{10}{30}{\decibel}.

We performed separation and dereverberation for \numlist{2;3;4} sources for 333 simulated mixtures.
The sampling frequency was \SI{16}{\kHz}, and the STFT frame size \num{1024} (\SI{64}{\milli\second}) is with three-quarter overlap.
We used a Hann window for analysis and the optimally matching window for synthesis.
The proposed methods were compared with ILRMA-T-IP~\cite{Ikeshita:2019:WASPAA}, ILRMA-IP~\cite{Kitamura:2016:ASLP}, and ILRMA-ISS\@.
We also evaluated ILRMA-IP and ILRMA-ISS initialized by WPE~\cite{Nakatani:2010:ASLP}, that we call WPE+ILRMA-IP and WPE+ILRMA-ISS, respectively.
For all ILRMA-T-based methods; ILRMA-T-ISS-JOINT, ILRMA-T-ISS-SEQ, ILRMA-T-IP,
we set the tap length $\NumTap$ to \num{5},
the delay parameter $\Delay$ to \num{2},
the initial DR and BSS filter $\{\derevMatrix _{\IdxFreq}\}_{\IdxFreq}$ to $\begin{bmatrix}\bm{I} _{\NumSource} & \bm{0} _{\NumSource\NumTap} \end{bmatrix}$,
respectively.
For all ILRMA-based methods; ILRMA-ISS and ILRMA-IP,
we set the initial BSS filter $\{\derevMatrix _{\IdxFreq}\}_{\IdxFreq}$ to the identity matrix.
For all methods, we set
the number of iterations to \num{100},
the number of NMF bases $\NumBasis$ to \num{2},
initial value of $\{\NMFbasis _{\IdxSource,\IdxFreq,\IdxBasis}\}_{\IdxSource,\IdxFreq,\IdxBasis}$ to \num{1},
and
initial value of $\{\NMFactivation _{\IdxSource,\IdxBasis,\IdxTime}\}_{\IdxSource,\IdxBasis,\IdxTime}$ to a random number uniformly distributed over $\interval[open right]{0.1}{1}$, respectively.
After separation and dereverberation, the scale of the output was restored by projection back onto the first microphone~\cite{Murata:2001:Neurocomputing}.

\newcommand{\dSISDR}{$\Delta$SI-SDR}
\newcommand{\dSISIR}{$\Delta$SI-SIR}
\subsection{Results}
We measured
the scale-invariant signal-to-distortion ratio (SI-SDR) and the scale-invariant signal-to-interference ratio (SI-SIR)~\cite{LeRoux:2019:ICASSP},
% the scale-invariant signal-to-distortion ratio (SI-SIR) [\si{\decibel}],
the cepstrum distance (CD),
and
the speech-to-reverberation modulation energy ration (SRMR).
We define \dSISDR{} and \dSISIR{} as the difference of SI-SDR and SI-SIR, respectively, between before and after the processing.

\Cref{fig:box} shows the separation performance after 100~iterations of each algorithm.
As a whole, the proposed ILRMA-T-based methods significantly outperformed the conventional ILRMA-based methods.
Also, they can slightly improve performance compared with WPE+ILRMA-IP and WEP+ILRMA-ISS\@.
\dSISDR{} and \dSISIR{} of ILRMA-T-ISS are slightly less than that of ILRMA-T-IP but achieve comparable performance in less time, as described below.
We can find that dereverberation improves the separation performance.
The proposed ILRMA-T-ISS-JOINT and ILRMA-T-ISS-SEQ can achieve comparable performance to ILRMA-T-IP\@.
This result is consistent with the reported difference between IP and ISS-based methods for BSS~\cite{Scheibler:2020:ICASSP}.

\Cref{fig:sdr:line} shows the comparison of convergence speed.
The total runtime of the proposed ILRMA-T-ISS-SEQ is about the same as that of ILRMA-T-IP, where $\NumSource = 2$.
On the other hand, it is much less where $\NumSource = 3, 4$.
The convergence speed of the proposed ILRMA-T-ISS is slightly slower than ILRMA-T-IP but the final performance is the same.
WPE+ILRMA-ISS and WPE+ILRMA-IP seem to converge the fastest, but the WPE initialization time was not included in the figure.

\section{Conclusion}
In this paper, we proposed a joint optimization technique for source separation and dereverberation based on ILRMA-T with ISS\@.
We use this technique to derive two new algorithms.
ILRMA-T-ISS-JOINT performs a sequence of ISS updates corresponding to the separation part of the algorithm, followed by a joint update corresponding to the parameters of the dereverberation.
Interestingly, this can be seen as a combination of the ISS and WPE updates applied alternatingly.
This form of the algorithm reduces the number of matrix inversion to just one per iteration and source.
ILRMA-T-ISS-SEQ gets rid of inversion altogether by updating all parameters via ISS-style rules.
This algorithm is very simple and does not need fancy linear algebra libraries.
It is a very good candidate for processing in practical edge or embedded systems.
Experimental results showed that while conceptually simpler, the proposed method performs just as well on a challenging dataset of noisy reverberant speech mixtures.
In future work, we intend to push the method towards real-time applicability, and explore advantages provided by extra microphones, the so-called overdetermined case~\cite{Togami:2020:APSIPA}.

\bibliographystyle{IEEEtran}
\bibliography{IEEEabrv,styles/conf_abrv_cap,ref}

\end{document}